\documentclass[twocolumn,superscriptaddress,PRL,reprint]{revtex4-1}

\usepackage{amsmath}
\usepackage{amssymb}
\usepackage{graphicx}
\usepackage{color}
\usepackage{bm}
\usepackage{enumerate}

\begin{document}

\title{Quantum vortex core and missing pseudogap in the multi-band BCS-BEC-crossover superconductor FeSe}

\author{T.~Hanaguri}
\email{hanaguri@riken.jp}
\affiliation{RIKEN Center for Emergent Matter Science, Wako, Saitama 351-0198, Japan}

\author{S.~Kasahara}
\affiliation{Department of Physics, Kyoto University, Kyoto 606-8502, Japan}

\author{J.~B\"oker}
\affiliation{Institut f\"ur Theoretische Physik III, Ruhr-Universit\"at Bochum, D-44801 Bochum, Germany}

\author{I.~Eremin}
\affiliation{Institut f\"ur Theoretische Physik III, Ruhr-Universit\"at Bochum, D-44801 Bochum, Germany}

\author{T.~Shibauchi}
\affiliation{Department of Advanced Materials Science, University of Tokyo, Chiba 277-8561, Japan}

\author{Y.~Matsuda}
\affiliation{Department of Physics, Kyoto University, Kyoto 606-8502, Japan}

\date{\today}

\begin{abstract}
{
FeSe is argued as a superconductor in the Bardeen-Cooper-Schrieffer Bose-Einstein-condensation crossover regime where the superconducting-gap size and the superconducting transition temperature $T_{\mathrm{c}}$ are comparable to the Fermi energy.
In this regime, vortex bound states should be well quantized and the preformed pairs above $T_{\mathrm{c}}$ may yield a pseudogap in the quasiparticle-excitation spectrum.
We performed spectroscopic-imaging scanning tunneling microscopy to search for these features.
We found Friedel-like oscillations near the vortex, which manifest the quantized levels, whereas the pseudogap was not detected.
These apparently conflicting observations may be related to the multi-band nature of FeSe.
}
\end{abstract}

\maketitle

Superfluidity in fermionic systems demands Cooper pairing by arbitrary strength of the pairing interaction.
In the weak coupling Bardeen-Cooper-Schrieffer (BCS) limit, Cooper-pair size is much larger than the inter-particle distance and the superfluidity occurs concurrently with the pair formation.
The other limit is the strong coupling Bose-Einstein-condensation (BEC) limit where the tightly-bound Cooper pairs are well described as weakly-interacting bosons.
The most intriguing situation is in between these two limits, BCS-BEC crossover regime, where the pair-formation instability and the phase-coherent superfluidity occur at different temperatures, $T_{\mathrm{ins}}$ and $T_{\mathrm{c}}$, respectively~\cite{Randeria2014ARCMP}.
A pseudogap due to preformed pairs develops between $T_{\mathrm{ins}}$ and $T_{\mathrm{c}}$ as detected in ultracold atomic fermions where the pairing interaction is tunable through the Feshbach resonance~\cite{Gaebler2010NatPhys}.

It is interesting to study the BCS-BEC-crossover regime in electronic counterparts, namely superconductors.
Since the pairing interaction is not tunable in a superconductor, we need a particular superconductor with the Cooper-pair size comparable to the inter-electron distance, namely $1/(k_{\mathrm{F}}\xi) \sim \Delta/E_{\mathrm{F}} \sim 1$.
Here, $k_{\mathrm{F}}$ is the Fermi momentum, $\xi$ is the coherence length, $\Delta$ is the superconducting gap, and $E_{\mathrm{F}}$ is the Fermi energy.
Almost all superconductors so far known are in the BCS limit where $\Delta/E_{\mathrm{F}} \ll 1$.

Recently, there appears accumulating evidence that an iron-based superconductor FeSe has large $\Delta/E_{\mathrm{F}}$.
FeSe is a compensated semimetal with hole bands at the Brillouin-zone center and electron bands at the zone corner.
Quantum oscillations~\cite{Terashima2014PRB,Watson2015PRB}, angle-resolved photomemission spectroscopy~\cite{Watson2015PRB}, and spectroscopic-imaging scanning tunneling microscopy (SI-STM)~\cite{Kasahara2014PNAS,Hanaguri2018SciAdv} revealed that Fermi surfaces and Fermi energies are very small, a few \% of the Brillouin zone and a few meV to a few tens of meV, respectively.
Superconducting gap is anisotropic and the maximum gap size is 1.5 $\sim$ 4~meV, depending on the band~\cite{Song2011Science,Sprau2017Science,Hanaguri2018SciAdv}.
Therefore, $\Delta/E_{\mathrm{F}} \sim 0.1$ or larger, placing FeSe in the BCS-BEC-crossover regime~\cite{Kasahara2014PNAS,Lee2017JCCMP}.

The signatures of the preformed pairs with unusually strong pairing fluctuations above $T_{\mathrm{c}}$ have been suggested by torque magnetometry~\cite{Kasahara2016NatCommun} and nuclear magnetic resonance~\cite{Shi2018JPSJ}.
However, mean-field-like BCS behaviors are reported in other experiments~\cite{Yang2017PRB,Takahashi2018arXiv}.
These controversial observations demand to examine the BCS-BEC-crossover signatures other than the superconducting fluctuations.
This is especially interesting as the BCS-BEC crossover in semimetals with hole and electron pockets was shown theoretically to have more peculiar features than that in the single band case~\cite{Chubukov2016PRB}.

To this end, we have performed two SI-STM experiments on FeSe.
First we examined the vortex core.
The vortex core consists of quantized bound states, whose number is $\sim E_{\mathrm{F}}/\Delta$~\cite{Caroli1964PhysLett}.
In the BCS limit, the vortex core is densely populated by a large number of levels that overlap and form a broad zero-energy peak in the local-density-of-states (LDOS) spectrum at the vortex center~\cite{Suderow2014SST}.
With increasing distance from the center, the zero-energy peak splits and continuously approaches to $\pm \Delta$.
By contrast, in the BCS-BEC-crossover regime, the vortex core accommodates only a few levels.
The lowest bound state is not at zero energy and the spatial evolution of the LDOS is no longer continuous because different discrete levels dominate depending on the distance from the vortex center~\cite{Hayashi1998PRL}.
The wave function of each bound state exhibits Friedel-like oscillations, and accordingly, the energy of the bound state shows stepwise change every $\pi/k_{\mathrm{F}}$.
We observed such Friedel-like oscillations, confirming the BCS-BEC crossover nature of FeSe.
The other experiment is the tunneling spectroscopy at elevated temperatures.
We found that superconducting gap closes just at $T_{\mathrm{c}}$ and no pseudogap was detected above it.
To address this apparent contradiction between two experiments, we analyze the BCS-BEC-crossover regime theoretically by taking the multi-band nature into account.
Using a two-band model with intra- and interband pairing channels and experimental parameters relevant for FeSe, we show that the pairing instability and the superconducting transition may occur simultaneously without a sizable pseudogap regime despite of large $\Delta/E_F$ in this system.

We used an ultra-high-vacuum low-temperature SI-STM system~\cite{Hanaguri2006JPCS} and performed experiments on \textit{in-situ} cleaved (001) surfaces of FeSe single crystals grown by vapor transport technique~\cite{Bohmer2013PRB}.
Chemically etched tungsten wires were used as scanning tips after cleaned by field evaporation and controlled indentation into clean Au(111) surfaces.
Tunneling conductance was measured by standard lock-in technique with modulation frequency of 617.3~Hz.

Figures~1(a)-1(e) show the tunneling-conductance maps, which represent the LDOS images, near a single vortex at different energies $E=eV$.
Here, $e$ is the elementary charge and $V$ is the sample bias voltage; negative and positive $E$ correspond to the filled and empty states, respectively.
The LDOS images are similar between filled and empty states, indicating that the observed patterns possess overall particle-hole symmetry as expected in Bogoliubov quasiparticles.
The images are strongly elongated along the orthorhombic $\mathbf{b}$ axis~\cite{Song2011Science} reflecting the nematic electronic state~\cite{Bohmer2018JPCM}.
(We adopt a coordinate system $|\mathbf{a}|<|\mathbf{b}|<|\mathbf{c}|$.)

\begin{figure}
\centering
\includegraphics[width=0.4\textwidth]{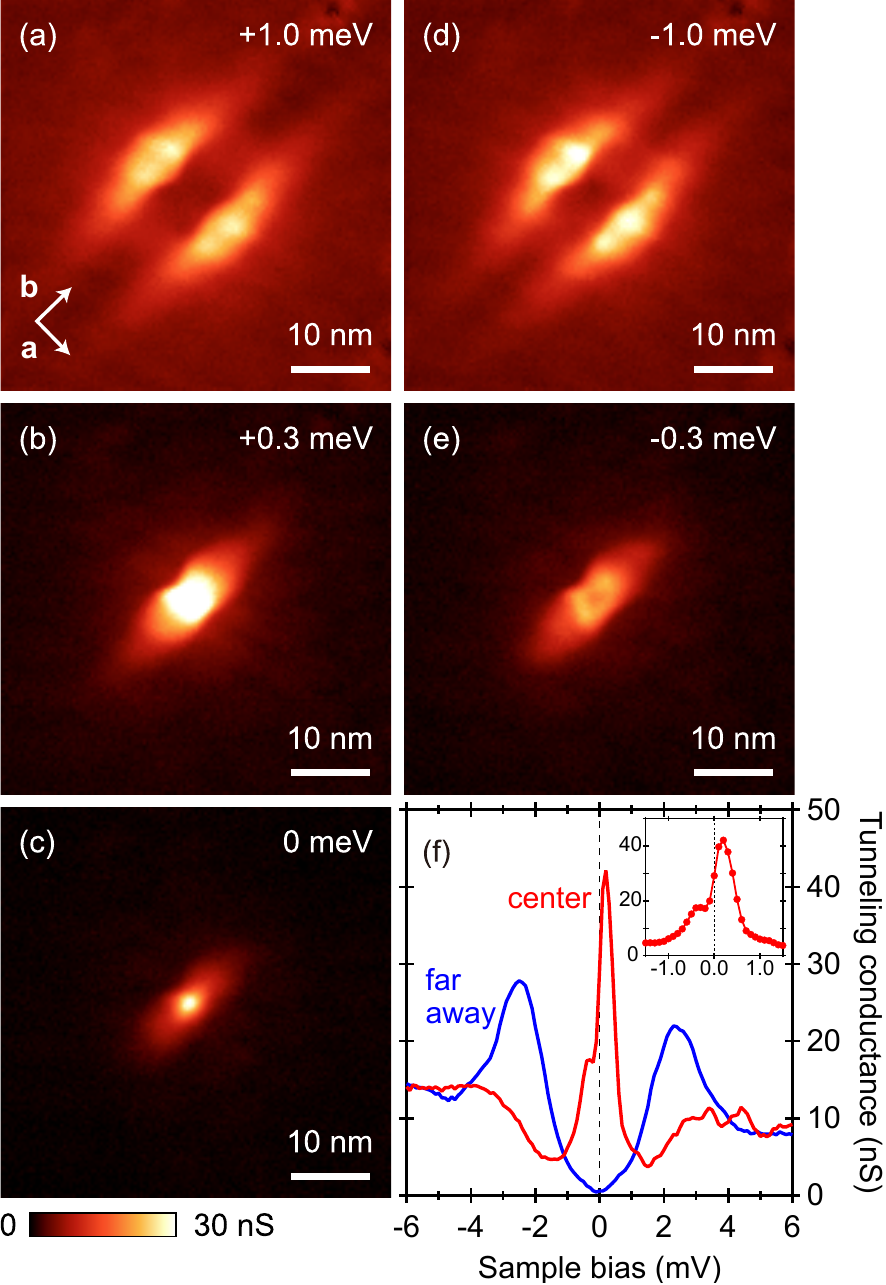}
\caption{
(a)-(e)~Tunneling-conductance maps showing a single vortex at different $E$.
Data were taken at 0.4~K in a magnetic field of 0.25~T along the $\mathbf{c}$ axis.
The tip was stabilized at tunneling current $I=100$~pA at $V=10$~mV.
Modulation amplitude for the lock-in detection $V_{\mathrm{mod}}=0.07$~mV$_{\mathrm{rms}}$.
(f)~Tunneling spectra taken at the vortex center (red) and away from vortices (blue).
$V_{\mathrm{mod}}=0.05$~mV$_{\mathrm{rms}}$.
Inset: Magnified spectrum at the vortex center.
}
\label{Fig1}
\end{figure}

Tunneling spectrum taken at the vortex center [Fig.~1(f)] possesses large spectral weight at low energies, which is characterized by two peaks at non-zero $E\sim \pm 0.2$~meV.
While the energies of these LDOS peaks are particle-hole symmetric, the spectral weight is much larger in the empty state than in the filled state.
Such an asymmetric spectral weights along with non-zero peak energies are indeed expected in the lowest bound state of the quantum vortex core~\cite{Hayashi1998PRL} in the BCS-BEC-crossover superconductor.
The asymmetry in the spectral weight is governed by the sign of the carrier~\cite{Hayashi1998PRL}, and the larger peak in the empty state means that the electron band is responsible for the peak formation.

\begin{figure*}
\centering
\includegraphics[width=0.7\textwidth]{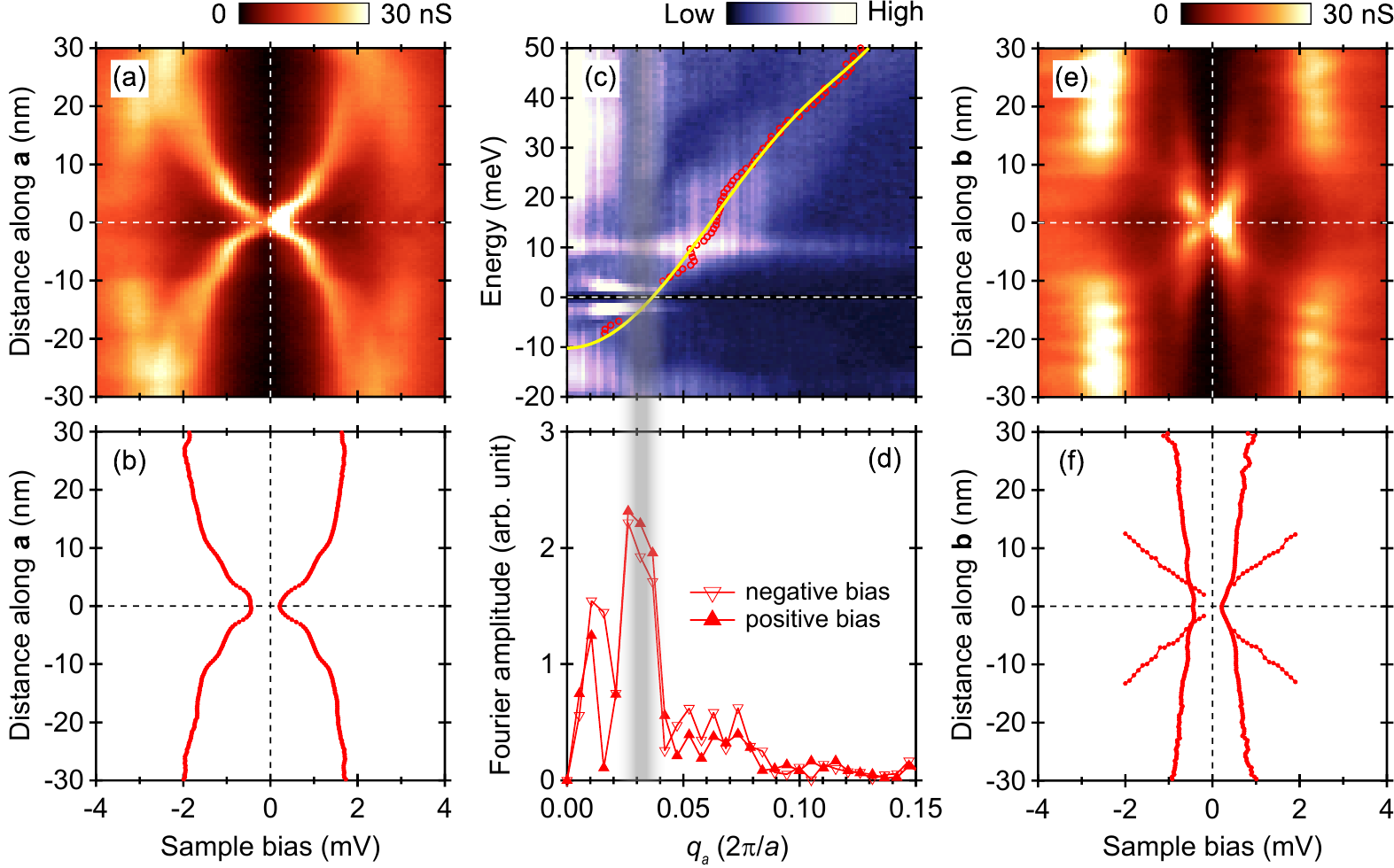}
\caption{
Spatial variations of the tunneling conductance~(a) and the bound-state energy~(b) along the $\mathbf{a}$ axis showing the oscillatory behaviors.
(c)~Energy-dependent line profile of the Fourier-transformed QPI pattern along the $\mathbf{q}_a$ direction.
(Data adopted from Ref.~\onlinecite{Hanaguri2018SciAdv}.)
Yellow solid line denotes the dispersion of the electron-band branch obtained by fitting the peak positions of the constant-energy line profile (red circles) to the polynomial function with even powers up to 6.
(d)~Fourier transforms from (b).
Smooth background was subtracted before the Fourier transformation.
As indicated by the vertical gray bar, the wave vector of the bound-state oscillation coincides with $\pi/k_{\mathrm{F}}$ obtained in (c).
Spatial variations of the tunneling conductance and the bound-state energy along the $\mathbf{b}$ axis are shown in (e) and (f), respectively.
}
\label{Fig2}
\end{figure*}

Next we searched for the Friedel-like oscillations by examining the line profiles of the conductance maps across the vortex center.
Figure~2(a) and 2(b) depict the evolutions of the LDOS spectra and LDOS-peak energies, respectively, along the $\mathbf{a}$ direction.
The LDOS-peak evolves with particle-hole symmetric stepwise changes.
If these are associated with the Friedel-like oscillations, the periodicity of the stepwise change should be $\pi/k_{\mathrm{F}}$.
We estimate $k_{\mathrm{F}}$ from the Fourier-transformed quasiparticle interference (QPI) pattern~\cite{Kasahara2014PNAS,Hanaguri2018SciAdv}.
Figure~2(c) shows the line profile from the Fourier-transformed QPI images along the scattering vector $\mathbf{q}_a \parallel \mathbf{a}$.
There is an electron branch that corresponds to the scattering vector along the minor axis of the elongated electron-band constant-energy contour~\cite{Hanaguri2018SciAdv}.
The magnitude of $\mathbf{q}_a$ at $E=0$ represents $\pi/k_{\mathrm{F}}$ that coincides well with the peak in the Fourier-transform of the spatial evolution of the LDOS-peak energy along the same direction [Fig.~2(d)].
This result indicates that the observed stepwise changes indeed represent the Friedel-like oscillations in the quantum vortex core, indicating that the electron band is in the BCS-BEC-crossover regime.
(See the Supplemental Material for details~\cite{SupplementalMaterial}.)

We also examined the line profile along the $\mathbf{b}$ direction where dispersions of the hole band are observed in the QPI patterns~\cite{Kasahara2014PNAS,Hanaguri2018SciAdv}.
As shown in Fig.~2(e) and 2(f), we identified two branches in the spatial evolutions of the LDOS-peak energies.
The origins of these branches are remained to be clarified.
Nevertheless, oscillatory behaviors are not observed in neither of vortex-bound-state branches [Fig.~2(f)], whereas the QPI modulations are seen in the intensities of the coherence peaks at $\sim \pm 2.5$~meV [Fig.~2(e)].
The absence of the vortex-bound-state Friedel-like oscillations along the $\mathbf{b}$ direction may be related to the three dimensionality and/or larger $E_{\mathrm{F}}$ (or $k_{\mathrm{F}}$) of the hole band.
(See the Supplemental Material for details~\cite{SupplementalMaterial}.)

\begin{figure}[b]
\centering
\includegraphics[width=0.47\textwidth]{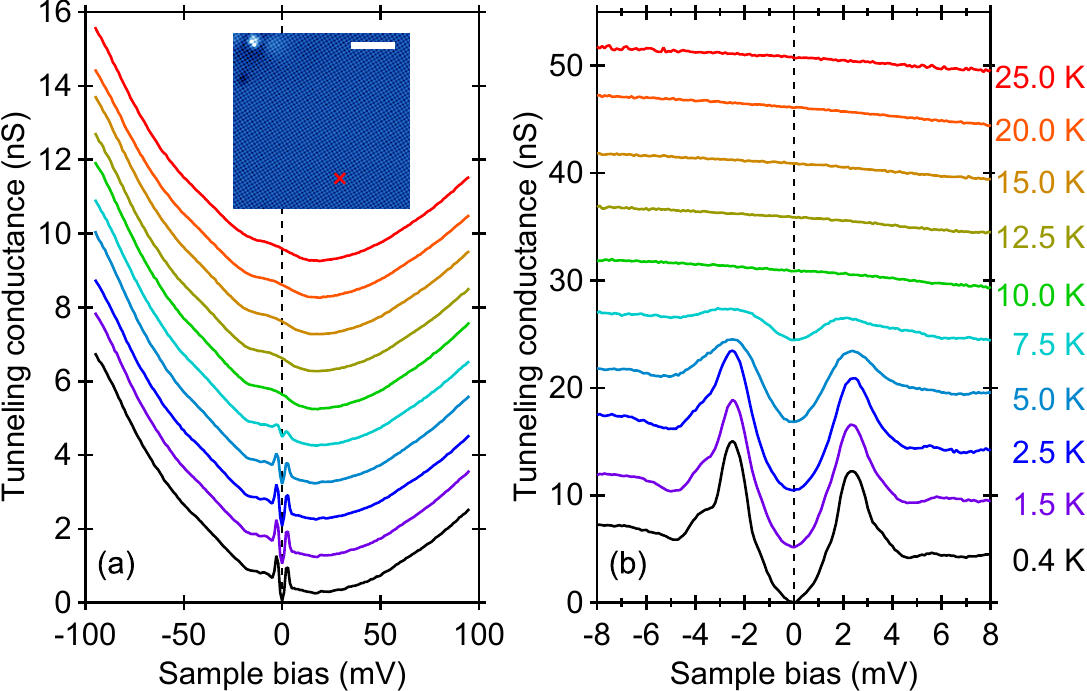}
\caption{
(a)~Tunneling spectra taken at different temperatures.
Measurement conditions were $I=100$~pA, $V=95$~mV and $V_{\mathrm{mod}}=0.7$~mV$_{\mathrm{rms}}$.
Each curve is shifted by 1~nS for clarity.
Inset: Topography showing the position (red cross) where the spectra were taken.
Scale bar corresponds to 5~nm.
(b)~Low-energy tunneling spectra taken at different temperatures.
Measurement conditions were $I=100$~pA, $V=20$~mV and $V_{\mathrm{mod}}=0.07$~mV$_{\mathrm{rms}}$.
Each curve is shifted by 5~nS for clarity.
}
\label{Fig3}
\end{figure}

Next we searched for the pseudogap above $T_{\mathrm{c}}$.
A pseudogap would be observed as a suppression in the tunneling spectrum at low energies.
Figure~3(a) shows the temperature dependence of the tunneling spectrum.
All the data were collected at the same atomic position that is $\sim 20$~nm away from the nearest defect [Fig.~3(a) inset].
The overall spectrum is U-shaped and a superconducting gap is identified at low energies.
High-resolution spectra [Fig.~3(b)] show that the superconducting gap disappears just above $T_{\mathrm{c}} \sim 9$~K but we should carefully examine whether the gap really closes or simply broadens due to thermal smearing.
A standard way to study this issue is fitting each spectrum to a model gap function smeared by the Fermi-Dirac function to extract the temperature-dependent gap amplitude.
However, FeSe possesses multiple anisotropic superconducting gaps~\cite{Song2011Science,Sprau2017Science}, preventing us from constructing a reliable model gap function.

\begin{figure*}
\centering
\includegraphics[width=0.8\textwidth]{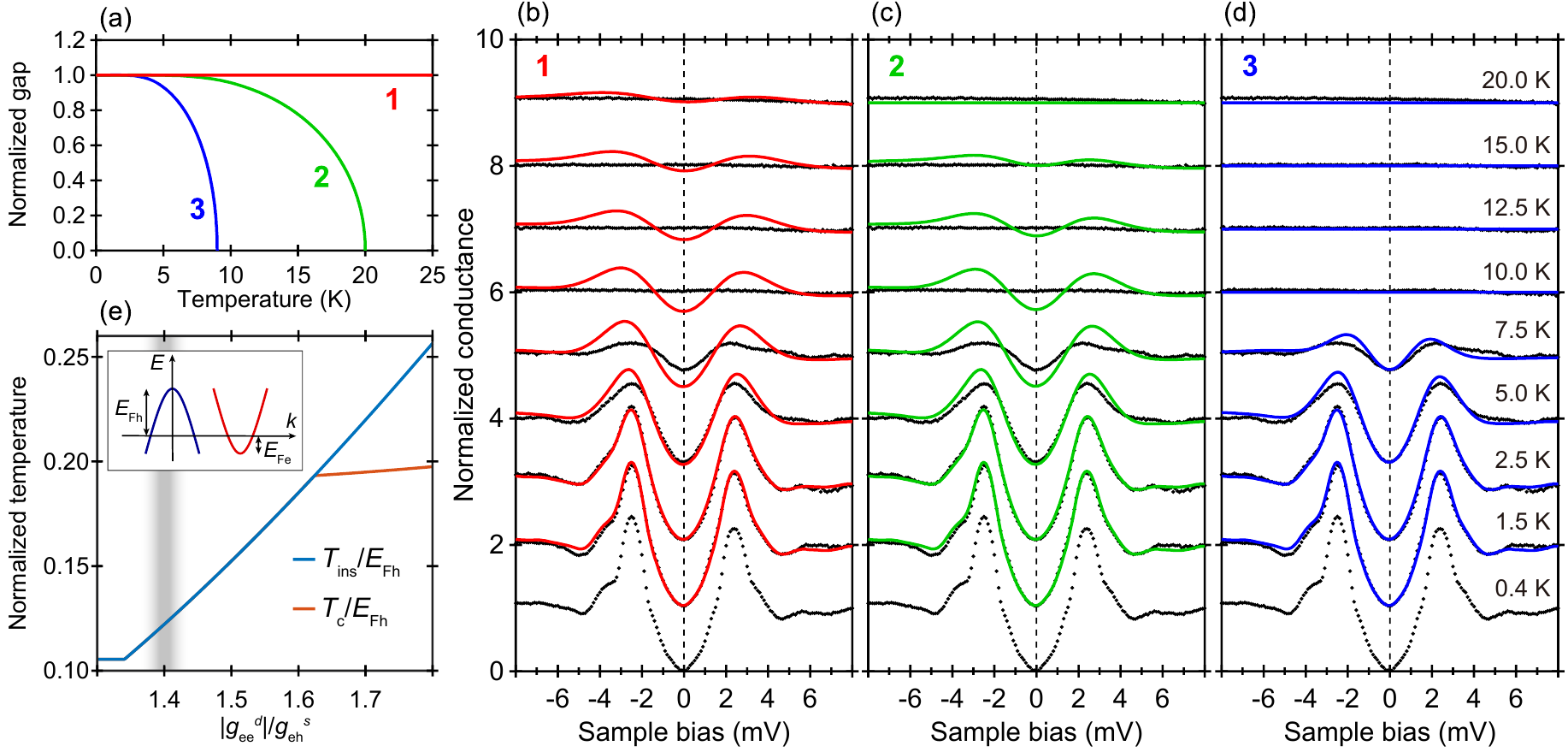}
\caption{
(a)~Temperature dependence of the gap amplitude assumed in the analysis.
(b)-(d)~Low-energy tunneling spectra normalized by the spectrum at 25~K (filled black circles) and their comparisons with the simulated spectra (solid lines).
For the simulated spectra in (b), (c), and (d), assumed $T_{\mathrm{v}}$'s are $\infty$, 20~K, and 9~K, respectively.
Each curve is shifted by 1 for clarity.
(e)~Calculated evolution of the $T_{\mathrm{ins}}$ and $T_{\mathrm{c}}$ with increasing intraband $d$-wave interaction.
The vertical gray bar refers to the parameter region relevant for FeSe.
Inset: The two-band electronic structure adopted in this model.
}
\label{Fig4}
\end{figure*}

Here we adopt the following alternative method.
We assume that all the gaps have the same BCS-like temperature dependence but the gap vanishing temperature $T_{\mathrm{v}}$ may be different from $T_{\mathrm{c}}$.
With this assumption, we can simulate the spectrum at finite temperature $T$ from that at $T=0$ by rescaling the energy axis with a scaling factor $\Delta_{\mathrm{BCS}}(T)/\Delta_{\mathrm{BCS}}(T=0)$, followed by the numerical thermal broadening.
Here, $\Delta_{\mathrm{BCS}}(T)$ denotes the $T$-dependence of the BCS gap~\cite{Noce1996NuovoCimento} that becomes zero above $T_{\mathrm{v}}$.
By comparing the spectrum simulated in this way with the observed one, we can judge whether the gap really closes at $T_{\mathrm{c}}$ or not.

We examined three different cases i.e. $T_{\mathrm{v}}=\infty$, $T_{\mathrm{v}}=20$~K, which corresponds to the onset of the strong superconducting fluctuations~\cite{Kasahara2016NatCommun}, and $T_{\mathrm{v}}=T_{\mathrm{c}}=9$~K [Fig.~4(a)].
We normalize the observed spectrum at each temperature by that at 25~K to remove the energy dependence of the background LDOS, and approximate the zero-temperature spectrum by the lowest-temperature (0.4~K) spectrum.
As shown in Fig.~4(b) and 4(c), the gap feature survives in the simulated spectra above $T_{\mathrm{c}}$ if we assume $T_{\mathrm{v}}=\infty$ and $T_{\mathrm{v}}=20$~K, indicating that actual $T_{\mathrm{v}}$ is lower than 20~K.
Indeed, the observed spectra are reasonably reproduced if $T_{\mathrm{v}}=T_{\mathrm{c}}=9$~K is assumed [Fig.~4(d)].
This means that the temperature range where the pseudogap remains is negligibly small, even if it exists.

The absence of the spectroscopic pseudogap is intriguing because we have observed the signature of large $\Delta/E_F$ in the same material, namely the Friedel-like oscillations in the vortex bound states.
We speculate that the multi-band character of FeSe and its compensated-metal structure may play an important role.
If the BCS-BEC-crossover superconductivity occurs in a single band system, the chemical potential may be shifted outside of the band edge because $\Delta \sim E_{\mathrm{F}}$.
However, if there are hole and electron bands, which nearly compensate each other, the chemical potential should be pinned at the original energy position~\cite{Chubukov2016PRB}.
Asymmetry between the electron and hole bands that should exist in a real material may alter the above simple picture.
The situation is further complicated by the fact that superconductivity in FeSe occurs below the structural (nematic) transition, which results in the superconducting order parameter being a mixture of the $s$- and $d$-wave symmetries.
Although it is believed that the dominant interaction in the $s$-wave channel is the interband scattering between hole and electron pockets, which supports the extended $s$-wave symmetry ($s^{\pm}$-wave), the dominant interaction in the $d$-wave channel most likely involves the scattering within the same type of the bands.
The splitting between $T_{\mathrm{ins}}$ and $T_{\mathrm{c}}$ in multi-band superconductors is more pronounced when dominant interaction occurs for the bands with similar character~\cite{Chubukov2016PRB}.
As both channels are mixed in FeSe it is interesting to study whether one would expect a pseudogap features in this material.

To address this question we consider a simplified interacting two-band model with orthorhombic distortions in two dimensions with hole and electron pockets with small Fermi energies [Fig.~4(e) inset].
We assume superconductivity due to repulsive interactions.
While the dominant interband interaction favors pairing in the $s^{\pm}$-wave symmetry, the $d$-wave projected intraband interaction yields attraction in the $d$-wave channel.
We obtained $T_{\mathrm{ins}}$ from the solution of the linearized mean-field gap equations, whereas $T_{\mathrm{c}}$ is estimated from superfluid stiffness.
Details of the calculations are given in the Supplemental Material~\cite{SupplementalMaterial}.

To find whether FeSe may have a sizable pseudogap, we analyzed the evolution of $T_{\mathrm{ins}}$ and $T_{\mathrm{c}}$ for various ratios of the dimensionless coupling constant in the $d$- and $s^{\pm}$-wave channels, $g^d_{\mathrm{ee}}$ and $g^s_{\mathrm{eh}}$, respectively [Fig.~4(e)].
From the experiment, we assumed $E_{\mathrm{F_h}} \sim 20$~meV, and $E_{\mathrm{F_e}} \sim 10$~meV~\cite{Kasahara2014PNAS,Hanaguri2018SciAdv}.
(Hereafter, subscripts e and h stand for electron and hole bands, respectively.)
Note that if interband $s^{\pm}$-wave dominates, $T_{\mathrm{ins}}$ and $T_{\mathrm{c}}$ do not split as long as one of the chemical potentials does not become negative~\cite{Chubukov2016PRB}.
As the intraband $d$-wave interaction grows, $T_{\mathrm{ins}}$ and $T_{\mathrm{c}}$ split for the regions where the superconducting gap has a dominant $d$-wave symmetry.
This, however, does not agree with the experiments in FeSe where the gaps are found to be nearly equal mixture of $s$- and $d$-wave harmonics~\cite{Sprau2017Science}.
By examining the evolution of the gaps as a function of $|g^d_{\mathrm{ee}}|/g^s_{\mathrm{eh}}$, we found that $s$- and $d$-wave gaps become comparable at $|g^d_{\mathrm{ee}}|/g^s_{\mathrm{eh}} \sim 1.4$~\cite{SupplementalMaterial} where $T_{\mathrm{ins}}$ and $T_{\mathrm{c}}$ do not split [Fig.~4(e)].
This indicates that despite of large $\Delta/E_{\mathrm{F}}$, multi-band nature prevents FeSe from pseudogap formation.
It is interesting to notice that in the pure $d$-wave case the splitting occurs more likely, which could be the situation in FeSe$_{1-x}$S$_x$ outside of the nematic phase~\cite{Hosoi2016PNAS,Hanaguri2018SciAdv}.
Nevertheless large superconducting fluctuations~\cite{Kasahara2016NatCommun,Shi2018JPSJ} may still be possible if there is an incipient band such as the inner hole band of FeSe~\cite{Watson2015PRB,Hanaguri2018SciAdv}, which contributes to the fluctuations but does not affect the LDOS near $E_{\mathrm{F}}$~\cite{Chubukov2016PRB}.

Our results clearly indicate that BCS-BEC-crossover in a multi-band system possesses a unique feature that is absent in a single band system.
A multi-band situation may be difficult to be achieved in ultracold atomic fermions, and thus FeSe offers a unique playground to search for as-yet-unknown novel phenomena in strongly interacting fermions.

\begin{acknowledgments}
We thank Y. Kato, K. Machida and Y. Nagai for valuable comments.
This work has been supported by Grants-in-Aid for Scientific Research from the Ministry of Education, Culture, Sports, Science and Technology of Japan (Grant Nos. 15H02106, 15H05852, 16H04024, 18H01177, 18H05227).
J.B. and I.E. were supported by the joint DFG-ANR Project (No.ER463/8-1) and DAAD PPP USA Grant No.N57316180.
\end{acknowledgments}

\end{document}